\documentclass[aps,preprint,superscriptaddress]{revtex4}
\usepackage{revsymb}
\usepackage{amsmath}
\usepackage{graphicx}
\newcommand{\be}{\begin{equation}}
\newcommand{\ee}{\end{equation}}
\newcommand{\ben}{\begin{eqnarray}}
\newcommand{\een}{\end{eqnarray}}

\begin{document}
\title{Statefinder parameters for interacting dark energy}
\author{Winfried Zimdahl\footnote{Electronic Mail-address:
zimdahl@thp.uni-koeln.de}}
\affiliation{Institut f\"{u}r Theoretische Physik, Universit\"{a}t zu K\"{o}ln,
50937 K\"{o}ln, Germany} 
\author{Diego Pav\'{o}n\footnote{Electronic Mail-address:  
diego.pavon@uab.es}}
\affiliation{Departamento de F\'{\i}sica Universidad Aut\'{o}noma 
de Barcelona, 08193 Bellaterra (Barcelona), Spain}
\begin{abstract}
We argue that the recently introduced ``statefinder parameters"
(Sahni et al., JETP Lett. {\bf 77}, 201 (2003)), that include the
third derivative of the cosmic scale factor, are useful tools to
characterize interacting quintessence models. We specify the
statefinder parameters for two classes of models that solve, or at
least alleviate, the coincidence problem.
\end{abstract}  
\maketitle 
\section{Introduction}
In the last years the conviction that our Universe is undergoing a
phase of accelerated expansion has gained further ground among
cosmologists \cite{sdss} albeit the nature of the cosmic substratum
(usually called dark energy) behind this acceleration remains as
elusive as ever \cite{paris}.  While several candidates have been
proposed \cite{review} there is no agreement on which of them should
be considered as the favored one. By all accounts, much more
observational input is needed before this point might 
be settled.

The acceleration is evaluated by the deceleration parameter
$q = - \ddot{a}/(aH^{2})$, where $a(t)$ stands for the scale factor of 
the Friedmann-Lema\^{\i}tre-Robertson--Walker (FLRW) metric and $H = \dot{a}/a$
for the Hubble parameter. As mentioned above, the present state
of the Universe seems to be characterized by a negative $q$, but 
it is hard to determine its value observationally. Therefore,
since many models predict acceleration some further information
should be welcome.

Among others, there are  cosmological models whose evolution is
dominated by interacting components -say, dark matter and dark
energy.  Models in which the main energy components do not evolve
separately but interact with each other bear a special interest
since they may alleviate or even solve the ``coincidence problem"
that afflicts many approaches to late acceleration
\cite{afflicts}. This problem can be summarily stated as ``why
now?", that is to say: ``Why the energy densities of the two main
components happen to be of the same order today?" In this paper we
focus on a Universe filled with two components, namely,
non-relativistic matter (subscript $m$) with negligible pressure,
i.e., $p_m \ll \rho_m$ and dark energy (subscript $x$) -with
equation of state $p_{x} = w \rho_{x}$ where $ w < 0$ -, such that
the latter decays into the former component according to
\\
\ben
\dot{\rho}_m & + & 3H \rho_m = Q \, , \nonumber\\
\dot{\rho}_x & + & 3H(1 + w) \rho_x = -Q \, ,
\label{conserv}
\een
\\
where $Q \geq 0$ measures the strength of the interaction. For later
convenience we will write it as $Q = - 3 \Pi H$ where the new
quantity $\Pi$ has the dimension of a pressure.

The Einstein field equations for spatially flat FLRW
cosmologies are
\\
\ben
H^{2} &=& \frac{8\, \pi G}{3} \rho \ ,  \\ 
\dot{H} & = & - \frac{8\, \pi G}{2} (\rho + p_x) \, ,
\label{efe}
\een
\\
where $\rho = \rho_m + \rho_x$ is the total energy density, and we
have set $c= 1$.
The quantity $\dot H$ is related to the deceleration parameter $q$ by
$q =- 1-(\dot H/H^2) = (1 + 3 w \Omega_{x})/2 $, where  $\Omega_{x} \equiv
8\, \pi G \rho_x/(3H^{2})$.
It is obvious, that the deceleration parameter does not depend on whether or
not both components are interacting. 
However, differentiating $\dot H$ again,
we obtain
\\
\be 
\frac{\ddot{H}}{H^{3}} = \frac{9}{2}\left(1 + \frac{p_x}{\rho}
\right) + \frac{9}{2}\left[w\left(1 + w \right)\frac{\rho_x}{\rho}
- w \frac{\Pi}{\rho} - \frac{\dot w}{3H}\frac{\rho_x}{\rho}
\right]\ . 
\label{doubledotH} 
\ee
\\
At variance with  $H$ and $\dot H$, the second derivative $\ddot H$
does depend on the interaction between the components.  Consequently,
to discriminate between models with different interactions or between
interacting and non-interacting models it is desirable to
characterize the cosmological dynamics additionally by parameters
that depend on $\ddot {H}$.
\noindent
Recently, Sahni et al. \cite{sahni} and Alam et al. \cite{alam} have
introduced a pair of new cosmological parameters (the so-called
``statefinder parameters") that seem to be promising candidates 
for this purpose.  These are:
\\
\begin{equation} 
r = \frac{\dddot{a}}{aH^{3}}, \qquad  s = \frac{r-1}
{3(q -\textstyle{1\over{2}})} \, . 
\label{statef} 
\end{equation}
\\
In the present context of interacting fluids they take the form
\\
\be r = 1+ \frac{9}{2}\frac{w}{1+ \kappa} \left[1 + w
-\frac{\Pi}{\rho _x} - \frac{\dot{w}}{3wH} \right] \ , 
\label{r2}
\ee
\\
where $\kappa \equiv \rho_m /\rho_x$, and
\\
\be
s = 1+ w -\frac{\Pi}{\rho_{x}} - \frac{\dot{w}}{3wH}\, . 
\label{s2} 
\ee
\\
For non-interacting models i.e., for $\Pi =0$, these parameters
reduce to the expressions studied in  \cite{sahni} and
\cite{alam}. The target of this short communication is to
illustrate how the statefinder parameters may be of help when
exploring cosmological models whose evolution is dominated by
interacting components. While we focus ourselves on interacting
cosmologies, we mention that the third derivative of the scale
factor is generally necessary to characterize any variation in the
overall equation of state of the cosmic medium \cite{visser}. This
becomes obvious from the general relation \cite{alam}
\\
\be
r - 1 = \frac{9}{2}\frac{\rho + P}{P} \frac{\dot{P}}{\dot
\rho} \ , \label{}
\ee
\\
where $P$ is the total pressure of the cosmic medium which in our
case reduces to $P \approx p_x$. Since
\\
\be
\left(\frac{P}{\rho}\right)^{\displaystyle \cdot} = \frac{\dot
\rho}{\rho} \left[\frac{\dot{P}}{\dot{\rho}} -
\frac{P}{\rho}\right]\ , \label{}
\ee
\\
it is evident, that an interaction term in $\dot{P}\approx
\dot{p}_x = \dot{w}\rho_x + w \dot{\rho}_x$ according to (\ref{conserv})
will additionally change the time dependence of the overall
equation of state parameter $P/\rho$.

Interacting models allow a dynamical approach to the coincidence
problem. The central quantity here is the density ratio $\kappa$ 
introduced beneath 
Eq. (\ref{r2}). This parameter should be a constant of
the order of unity at late times for the coincidence problem to be
strictly solved. At least it should, however, vary slowly over a
time of the order of $H^{-1}$. The ratio $\kappa$ is governed by
the evolution equation (cf. Eqs. (\ref{conserv}))
\\
\be 
\dot{\kappa} = - 3H \left[\left( \frac{\rho_{x} +
\rho_{m}} {\rho_{m} \, \rho_{x}}\right) \Pi - w \right] \kappa \, .
\label{evolr} 
\ee
\\
Below we study the statefinder parameters for different solutions
of this equation, corresponding to two broad classes
of matter--quintessence interacting models.

\section{Scaling solutions}
In a recent paper \cite{scaling} the authors showed that scaling
solutions, i.e., solutions of the form $\rho_m/\rho_x \propto
a^{-\xi}$, where $\xi$ denotes a constant parameter in the range
$[0,3]$, can be obtained when the dark energy component decays
into the pressureless matter fluid  -Eqs. (\ref{conserv}). These
solutions are interesting because they alleviate the coincidence
problem \cite{dalal}.  Indeed, a model with $\xi = 3$ amounts to
the $\Lambda$CDM model with $w = -1$ and $\Pi =0$. For the
opposite extreme value $\xi = 0$ the Universe dynamics admits a
stable, stationary solution $\kappa = $ constant, thereby no
coincidence problem arises \cite{ZPC}. Hence, the deviation of the
parameter $\xi$ from $\xi = 0$ quantifies the severity of the
problem whereby any solution that differs from $\xi = -3 w$
represents a testable, non--standard cosmological model and any
solution with $\xi < 3$ renders the coincidence problem less
acute. In that scheme, with $w = $ constant, it can be shown  that
the interactions which produce scaling solutions are given by
\\
\be \frac{\Pi}{\rho_{x}} = \left(w + \frac{\xi}{3}\right) \,
\frac{\kappa_{0} (1+z)^{\xi}}{1 + \kappa_{0} (1+z)^{\xi}}\  , \ee
\\
where $\kappa_0$ denotes the present energy density ratio and $z =
(a_{0}/a) - 1$ is the redshift. Inserting this expression in Eqs.
(\ref{r2}) and (\ref{s2}) we get for the statefinder parameters
\\
\begin{equation}
r =  1 + \frac{9}{2} \frac{w}{1 + \kappa_{0} (1+z)^{\xi}}
\left[1 + w - \left( w + \frac{\xi}{3} \right) \frac{\kappa_{0}
(1+z)^{\xi}} {1 + \kappa_{0} (1+z)^{\xi}}\right] \, ,
\label{r3}
\end{equation}
\\
and
\be
s = 1 + w -  \left( w + \frac{\xi}{3} \right)
 \frac{\kappa_{0} (1+z)^{\xi}}{1 + \kappa_{0} (1+z)^{\xi}}.
\label{s3}
\ee
\\
Figure 1 depicts the function $r(s)$ for different values of $\xi$.
The lower the value of $\xi$, the lower the value of the 
corresponding curve in the $s$-$r$ plane and the less acute results 
the coincidence problem. Note that qualitatively these curves remain
similar to those of non-interacting models (see Fig.1 of
Ref.\cite{sahni}). For the sake of comparsion, we note 
that the $\Lambda$CDM model ($\Pi = 0$, $w = -1$) 
corresponds to the point (not shown in the figure) 
$s = 0$, $r = 1$.

\begin{figure}[tbp]
\includegraphics*[scale=0.5]{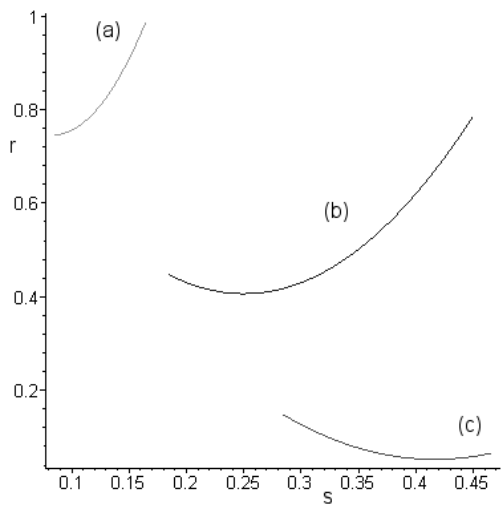}
\caption{Selected curves $r(s)$ in the redshift interval $[0,6]$ (from left 
to right) with $w=-0.95$ and $\kappa_{0} = 3/7$ for three different
values of the parameter $\xi$, {\em viz} (a) $2.5$; (b) $1.5$; (c) $0.5$.}
\label{fig1}
\end{figure}

\noindent
From an observational point of view and for discriminating between different 
models, the current values $r_0$ and $s_0$ of the statefinder parameters 
are of particular relevance. 
For the scaling models we have 
\\
\begin{equation}
r_0 = 1 + \frac{9}{2} \frac{w}{1 + \kappa_0}s_0 \, ,
\qquad {\rm and }\qquad
s_0 = 1+ w -  \left(w + \frac{\xi}{3}\right)
\frac{\kappa_0}{1 + \kappa_0} \, .
\label{rsscaling} 
\end{equation}
\\
Realizing that 
\\
\be 
q_0  = \frac{1}{2}
\frac{1 + \kappa_0 + 3w}{1 + \kappa_0}  \ , 
\label{} 
\ee
\\
and introducing 
\\
\be 
q_{0\Lambda}\equiv q_0\left(w = -1\right)  = -\frac{1}{2}
\frac{2 - \kappa_0}{1 + \kappa_0}  
\Leftrightarrow 
\frac{3}{2}\frac{\kappa_0}{1 + \kappa_0} = 1 + q_{0\Lambda}\ , 
\label{} 
\ee
\\
we may classify the different models through their dependence 
$s_0(q_0)$ which reads 
\\
\be 
s_0 = \frac{2}{3}\left[
q_0
- q_{0\Lambda}
+ \left(
\frac{\xi}{3} - 1
\right)\left(
1 + q_{0\Lambda}
\right)\right]\ .
\label{reads} 
\ee
\\
The first part in the bracket on the right hand side describes the deviation 
from $w = - 1$, the second part accounts for the deviations from 
the $\Lambda$CDM scaling $\xi = 3$. 
For models with $w = - 1$, e.g., which all have the same deceleration 
parameter 
$q_0 = q_{0\Lambda}$, we have 
$s_0 = \frac{2}{3}\left(\frac{\xi}{3} - 1 \right)
\left(1 + q_{0\Lambda}\right)$. 
Of course, $\xi = 3$ corresponds to the $\Lambda$CDM model 
with $s_0 = 0$. Assuming $\kappa_0 = 3/7$, 
a scaling $\xi = 1$ results in $s_0\left(\xi = 1\right) = - 0.2$, 
while the stationary solution $\xi = 0$ has 
$s_0\left(\xi = 0\right) = - 0.3$. 
Similar considerations hold for other values of $w$. 
Thus, the parameter $s_0$ is able to discriminate between  different 
scaling models, characterized by the same deceleration parameter.

\subsection{Luminosity distance}
It is interesting to see how the statefinder parameters enter the
expression for the luminosity distance.  Up to second order in the
redshift the Hubble rate is
\\
\be 
H\left(z\right) =H_0 + \left( \frac{\mbox {d} H}{\mbox {d}z}
\right)_{z=0} z + \frac{1}{2}  \left(\frac{\mbox{d}^2 H}
{\mbox{d} z ^2 } \right)_{z = 0} z^2 + ...\, . 
\label{hdz1} 
\ee
\\
By virtue of 
\\
\be 
\frac{\mbox {d} H}{\mbox {d}z} = \frac{q + 1}{z + 1}H 
\qquad \mbox{and} \qquad
\frac{\mbox{d}^2 H}
{\mbox{d} z ^2 } 
= \frac{r-1 + 2\left(1+q\right) - \left(1+q\right)^2}
{\left(1+z\right)^2}\, H\ ,
\label{virtue}
\ee
\\ 
this can be written as 
\\
\be
H\left(z\right) =H_0 \left\{1 + \left(q_0 + 1\right)z +
\frac{1}{2}\left[r_0 - 1 + 2 \left(q_0 + 1\right) - 
\left(q_0 + 1\right)^2\right]z^2 + ...\right\}\,\ .
\label{hdz2}
\ee
\\
The luminosity distance
\be
d_L =
\left(1+z\right)\int \frac{\mbox{d}z}{H}\ ,
\label{dL1}
\ee
\\
becomes (cf. \cite{visser})
\be
d_L = H_0^{-1}z \left[1 +
\frac{1}{2}\left(1 -q_0 \right)z + \frac{1}{6}\left( 3\left(q_0 +
1\right)^2 - 5\left(q_0 + 1\right) + 1 - r_0 \right)z^2 +
...\right]\  .
\label{dL2}
\ee
\\
Since the interaction affects $r_0$ but neither $q_0$ nor $H_0$, it is
obvious that the luminosity distances of different interacting as
well as of interacting and non-interacting models manifests itself
only in third order in the redshift. For the scaling model leading 
to $\xi =1$, and $w = -1$ we have 
$q_{0} = -(2-\kappa_{0})/[2(1+\kappa_{0})]$ 
and $r_{0} = 1- [3\kappa_{0}/(1+\kappa_{0})^{2}]$, and we recover the 
previously obtained expression (see Eq. (38) of Ref. \cite{scaling})

\begin{equation}
d_{\rm L} \approx H_0^{-1}z \left[1 + \frac{1 + 
\frac{\kappa_0}{4}}{1 + \kappa_0}z - \frac{1}{8}\frac{\kappa_0}
{\left(1+\kappa_0\right)^2}\left(6 + \kappa_0\right)z^2\right] \ . 
\label{dL3}
\end{equation}

\noindent
For the $\Lambda$CDM model the factor $\left(6 + \kappa_{0}\right)$ 
occurring in last expression is replaced by  
$\left(10 + \kappa_{0}\right)$. On the other hand, 
for a given $w$ all the scaling models, including $\Lambda$CDM, are
degenerate with respect to the deceleration parameter. For all these
models the present value of $q$ is
\\
\begin{equation}
q_0 = \frac{1}{2} + \frac{3}{2}\frac{w}{1 + \kappa_0}\ .
\label{q0}
\end{equation}
\\
This expression also holds true for the asymptotically stable model 
of the next section. This demonstrates explicitly that discrimination
between interacting and non-interacting models or between different
interacting models requires the knowledge of the luminosity distance
up to the third order in the redshift. In other words, the
circumstance that the interaction is felt only by parameters
containing the third derivative of the scale factor corresponds to
the fact that the luminosity distance of interacting models is
affected only in the third order of the redshift.

\section{Asymptotically stable solutions}
For the special case that the interaction term is assumed to obey
$\Pi = -c^{2} \rho$ with $c^{2} = \mbox{constant} < 1$, the
evolution equation (\ref{evolr}) has two stationary solutions for
$w =$ constant, namely, $\kappa_{s}^{+}$ and 
$\kappa_{s}^{-} = 1/\kappa_{s}^{+}$ (with $\kappa_{s}^{+} > 1$), 
given by
\\
\be \kappa_{s}^{\pm} = - \left[1 + \frac{1}{2c^2} \left(w \mp
\sqrt{w(w +4c^{2})}\right) \right] , \label{rpm} \ee
\\
-see Ref. \cite{iqs} for details. It can be shown that whereas the 
solution $\kappa_{s}^{-}$ is stable the solution $\kappa_{s}^{+}$ 
is unstable. There exists a solution 
\\
\be
\kappa = \kappa_{s}^{-} \frac{1 + y \kappa_s^+}
{1 + y \kappa_s^-}
\ , \label{kappa} \ee
\\
with $y = (a_{eq}/a)^{\lambda}$ where $a_{eq}$ is the scale
factor at which the energy density of dark energy equals the
energy density of dark matter and 
$\lambda = -3 w (1- \kappa_{s}^{-})/(1+ \kappa_{s}^{+})$, 
according to which  $\kappa$ evolves from a
matter dominated phase ($\kappa_{s}^{+} >1$) for $a \ll a_{eq}$  to a 
dark energy dominated phase ($\kappa_{s}^{-} < 1$) for $a \gg a_{eq}$
as the Universe expands. In this case we have
\\
\be 
\frac{\Pi}{\rho_{x}} = - c^{2} (1+ \kappa_{s}^{-}) \frac{1 +
y_{o} (1+z)^{\lambda}} {1 + y_{o} (1+z)^{\lambda} \,
\kappa_{s}^{-}}\ ,
\label{piasympt} \ee
\\
with 
$y_{0} =\kappa_{s}^{+}(\kappa_{0}-\kappa_{s}^{-})/(\kappa_{s}^{+}-\kappa_{0})$.

Accordingly, we readily obtain  that the statefinder parameters of this
model reduce to
\\
\be
r = 1+ \frac{9}{2} w \left[\frac{1+w}{1+\kappa_{s}^{-}} \,
\frac{1 + y_{o} (1+z)^{\lambda} \kappa_{s}^{-}}{1+y_{o}
(1+z)^{\lambda}} + c^{2}\right]
\label{r4}
\ee
\\
and
\be
s = 1+w+c^{2}(1+\kappa_{s}^{-}) \frac{1 + y_{o}
(1+z)^{\lambda}} {1 + y_{o} (1+z)^{\lambda} \, \kappa_{s}^{-}} \, ,
\label{s4}
\ee
\\
where $c^{2} = - w \kappa_{s}^{-}/(1+\kappa_{s})^{2}$ is valid.

Figure 2 shows some graphs of the function $r(s)$ for different choices
of $\kappa_{s}^{-}$.  At variance with graphs of the scaling model, the
location of the curves in the plane $(s,r)$ is unrelated to the
alleviation or solution of the coincidence problem since 
$\kappa= \kappa_{s}^{-} =$ constant at late times for 
all the cases of this model.

\begin{figure}[tbp]
\includegraphics*[scale=0.5]{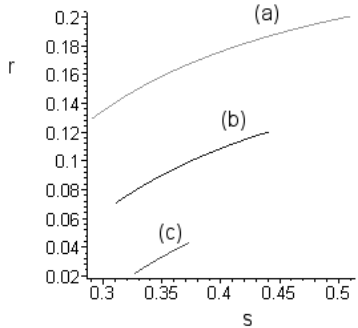}
\caption{Selected curves $r(s)$ in the redshift interval $[0,6]$
(from left to right) with $w = -0.95$ and $\kappa_{0} = 3/7$, for
different values of $\kappa_{s}^{-}$, {\it viz} (a) 0.3; (b) 0.35;
(c) 0.4.}
\label{fig2}
\end{figure}

For the present value of the parameter $s$ this model yields
\\ 
\be
s_0 = 1+w+c^{2}\left(1+\kappa_0 \right)\ , \quad {\rm where}\quad 
1+\kappa_0 = (1+\kappa_{s}^{-}) \frac{1 + y_{o}} 
{1 + y_{o}\, \kappa_{s}^{-}} \, .
\label{yields}
\ee
\\
A parallel study to the case of the scaling solutions leads to
\\
\be
s_0 = \frac{2}{3}\left[q_0 - q_{0\Lambda} + \frac{3}{2}c^2
\right]\left(1 + \kappa _0\right)
\, .
\label{parallel}
\ee
\\
Again, the first two terms in the bracket account for the difference to models 
with $w=-1$ for which $q_0 =q_0(w=-1)\equiv q_{0\Lambda}$. 
The $c^2$ term describes the impact of the interaction on the 
parameter $s_0$.

\section{Concluding remarks}
The statefinder parameters introduced in \cite{sahni} and \cite{alam}
are expected to be useful tools in testing interacting cosmologies
that solve or at least alleviate the coincidence problem which besets
many approaches to late acceleration. It is manifest that while the
deceleration parameter does not feel the interaction between the dark
energy and dark matter the statefinder pair $(r, s)$ does.

We hope that in some not distant future we will have at our disposal
observational techniques capable of  determining these parameters.
These are bound  to shed light on the nature of dark energy and dark
matter.

\section*{Acknowledgments}
This work was supported by the Deutsche Forschungsgemeinschaft and the
Spanish Ministry of Science and Technology under grants BFM 2000-C-03-01
and 2000-1322.

\end{document}